\documentclass[aps,prd,twocolumn,superscriptaddress,letterpaper]{revtex4}

\usepackage{dcolumn}
\usepackage{bm}
\usepackage{epsfig}
\usepackage{graphicx,psfrag}

\newcommand{\be}{\begin{equation}}
\newcommand{\ee}{\end{equation}}
\newcommand{\bear}{\begin{eqnarray}}
\newcommand{\eear}{\end{eqnarray}} \newcommand{\ba}{\begin{array}}
\newcommand{\ea}{\end{array}}

\def\beq{\begin{equation}}
\def\eeq{\end{equation}}
\def\beqa{\begin{eqnarray}}
\def\eeqan{\end{eqnarray}}

\newcommand\iden{\leavevmode\hbox{\small1\normalsize\kern-.33em1}}

\def\W3{W_H^3}

\begin{document}
\setcounter{footnote}{0}

\title{Detection of Exotic Massive Hadrons in Ultra High Energy
Cosmic Ray Telescopes}
\author{Ivone~F.~M.~Albuquerque}
\affiliation{Instituto de F\'isica, Universidade de S\~ao Paulo, 
S\~ao Paulo, Brazil and Fermi National Accelerator Laboratory, 
Batavia, Illinois, USA.}
\author{Washington~R.~Carvalho~Jr.}                     
\affiliation{Instituto de F\'isica, Universidade de S\~ao Paulo,
S\~ao Paulo, Brazil}

\vspace*{0.3cm}
\date{\today}


\begin{abstract}
We investigate the detection of exotic massive strongly interacting 
hadrons (uhecrons) in ultra high energy cosmic ray telescopes. 
The conclusion is that experiments such as the Pierre Auger Observatory 
have the potential to detect these particles. It is shown that
uhecron showers have clear distinctive features when compared to proton and
nuclear showers. The simulation of uhecron air showers, and its
detection and reconstruction by fluorescence telescopes is described. 
We determine
basic cuts in observables that will separate uhecrons from the cosmic
ray bulk, assuming this is composed by protons. If these are composed
by heavier nucleus the separation will be much improved. We also discuss
photon induced showers. The complementarity between uhecron detection in 
accelerator experiments is discussed.
\end{abstract}
\maketitle

\section{Introduction} \label{sec:intro} 
Ultra high energy cosmic ray (UHECR) observatories 
investigate the high energy end of the cosmic ray spectrum (above $\sim 10^{19}$~eV). 
Their results \cite{hrflx,auflx} are consistent with the
presence of the Greisen \cite{greisen} and Zatsepin and Kuzmin 
\cite{ZK} (GZK) feature. 

GZK showed that nucleons propagating through the Cosmic Microwave Background 
Radiation (CMB) will have their energy degraded. The main energy loss 
mechanism for cosmic rays above $\sim 5 \times 10^{19}$~eV is pion
photoproduction. In order to reach the Earth, nucleons have to 
be produced relatively near us, at a maximum distance of about 100 Mpc.
As a consequence, the cosmic ray energy spectrum should fall steeply
around $\sim 5 \times 10^{19}$~eV. This feature is known as the GZK cutoff.
Since there are events \cite{agasa,hires} detected beyond this cutoff, their origin, 
composition and sources became a puzzle and the existence of the GZK
cutoff was questioned.

Here we investigate the possibility of detecting exotic massive and strongly
interacting hadrons (uhecrons) in the Pierre Auger Observatory \cite{auger}.
Uhecrons were first proposed \cite{cfk} as a solution to the GZK \cite{greisen,ZK}
puzzle. Due to their greater mass, their threshold energy for pion photoproduction is
larger than for a proton. For this reason, 
an uhecron's energy degradation through the CMB is much smaller when compared 
to a proton \cite{cfk} and it can come from farther away. A thorough search
\cite{sommers} for the source of the highest energy cosmic ray ever detected (by the Fly 
Eye's collaboration \cite{flys}), pointed to a faraway ($z = 0.545$) quasar (3C147) as
one of the best candidates. Although a proton coming from this distance can not reach
the Earth, an uhecron can.

Uhecron candidates are found in extensions of the standard model of particle physics.
Heavier uhecrons (masses $> 50$~GeV) were excluded \cite{afk} as UHECR. Besides
other reasons, the  air showers they produce have their maximum 
too deep in the atmosphere. Among the surviving candidates are the
heavy gluino lightest supersymmetric particle (LSP) \cite{raby,rabycdf} and 
strongly interacting wimpless dark matter \cite{feng}. A search for the heavy gluino LSP 
using  CDF \cite{rabycdf} and LEP data \cite{hglep} constrained its mass to a 
25 to 35~GeV window. Here we show that the neutral mode of this particle can be detected 
by UHECR telescopes, and this mass window allows for discrimination from the bulk of UHECR 
assuming it is composed by protons or nucleus. 

Our investigation \cite{tesew} follows the uhecron
definition stated in \cite{afk}. It is an electrically neutral, strongly interacting 
heavy exotic hadron. The bulk of its mass is carried by a 
single constituent. This is surrounded by hadronic degrees of freedom, which
are responsible for the uhecron interaction.

We simulate uhecron induced air showers in a similar way as described in \cite{afk} and then 
the detection and event reconstruction by a fluorescence detector (FD) as 
described in \cite{fluor}. Proton and uhecron induced showers are compared and
their discriminating parameters are determined. As all UHECR simulations extrapolate
known physics at lower energies to much larger energies, it is important to note
that we compare uhecron to proton observables. In this
way, we reduce the bias introduced due to uncertainties in the extrapolation of interaction
models to high energies, since these uncertainties will affect both protons and uhecrons.

As a result, we show that uhecrons with masses below 50~GeV can be detected in
UHECR telescopes and discriminated against protons and nucleus. 

In the next section we describe our simulation of uhecron induced showers. Follows the
description of the FD detection and event reconstruction simulation. We then
describe the main uhecron induced shower features and compare them to proton and iron
induced showers. In section~\ref{sec:analy} we describe how to discriminate between
protons and uhecrons. In the following section we discuss
photon induced showers. The last section presents our conclusions.

\section{Uhecron Induced Shower Simulation}
\label{sec:uhesim}

When a UHECR impinges the Earth atmosphere, it generates a shower of particles.
As the shower develops, the number of particles increases until it reaches a 
maximum at a certain
point in the atmosphere (Xmax). At this maximum the energy of each particle is
low enough to be lost through ionization. The development of the shower as a function
of the atmospheric depth (longitudinal profile) depends on the primary cosmic ray
composition. The longitudinal profile integrated energy 
is proportional to the primary cosmic ray energy.

Air shower simulations include a particle cascade development integrated with 
an event generator. The later simulates the interactions between particles with air 
nucleus
while the shower development simulates the particle cascade versus atmospheric
depth. In our simulation we use the Air Shower Extended Simulations (AIRES) 
package (v2.8.4a) \cite{aires} with SIBYLL (v2.1) \cite{sibyll,sibnuc} as the event generator.

In order to simulate uhecron induced showers, we modified both AIRES and SIBYLL.
While modifications to AIRES are straight forward, and basically requires the inclusion
of a new particle in
the shower development, the modifications to the event generator are more complex.
We use the modifications described in details in \cite{afk}.

SIBYLL \cite{sibyll} models the interaction of a particle with an air nucleus as a
combination of a low energy hadron-hadron interaction and a model for the ``hard''
part of the cross section. It also models hadron-nucleus interactions \cite{sibnuc}.
The interactions that occur high in the atmosphere have very large center of
mass (CM) energies. SIBYLL extrapolates the known physics at much lower energies
to higher CM energies ($\vartheta(100$~TeV)) using the dual
parton model \cite{dual} superposed by minijet production.

In short, the main modifications to SIBYLL (described in more details in \cite{afk}) 
are as follows. The uhecron is represented as
a heavy single constituent ($Q$) surrounded by light hadronic degrees of 
freedom. Its interaction is simulated in the same way as hadron-hadron interactions,
which are represented \cite{sibyll} by production and fragmentation of QCD strings.
However,  uhecron interactions use harder structure and
fragmentation functions than the ones used for normal hadrons.
In analogy to the B meson, we describe the 
fraction of energy $z$ carried by $Q$, using the Peterson fragmentation function
\cite{pet,pdg}:

\beq
f_Q(z) \;=\; \frac{1}{z} \left[ 1 - \frac{1}{z} - \frac{\varepsilon_Q}{1 - z}
\right]^{-2}
\eeq
where $\varepsilon_Q$ is proportional to $\Lambda^2_{QCD} / m_Q^2$. 

The good agreement between this fragmentation function and data is described 
in \cite{aleph}.
This guarantees that most of the uhecron momentum is carried by the heavy constituent. 
The same function is used
for the structure function, which describe the fraction of energy of the hadron 
carried by $Q$.

As the light constituents are responsible for the interactions, we take the
uhecron-nucleon ($\sigma_{UN}$) cross section as the one for pion-nucleon interactions.
Other modifications are related to diffraction dissociation,
where the lower mass limit of the excited state was modified according to the
uhecron mass ($m_U$). Also the ``hard'' part of the cross section
with large momentum transfer, which is simulated as
minijet production, is turned off for uhecrons, since most of the momentum is carried
by $Q$ which does not interact. 

\begin{figure}[h]
\includegraphics[scale=0.35]{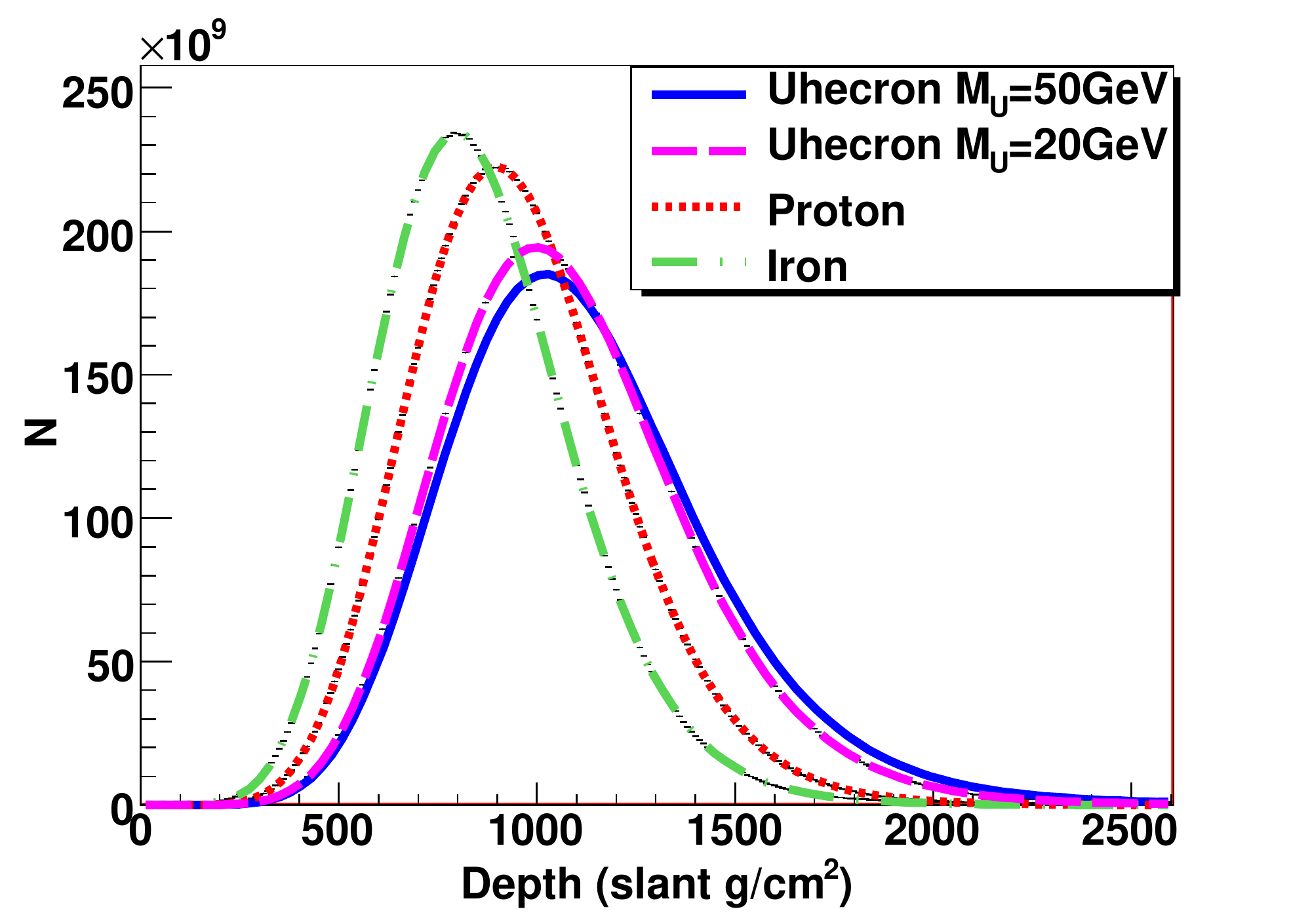}
\includegraphics[scale=0.35]{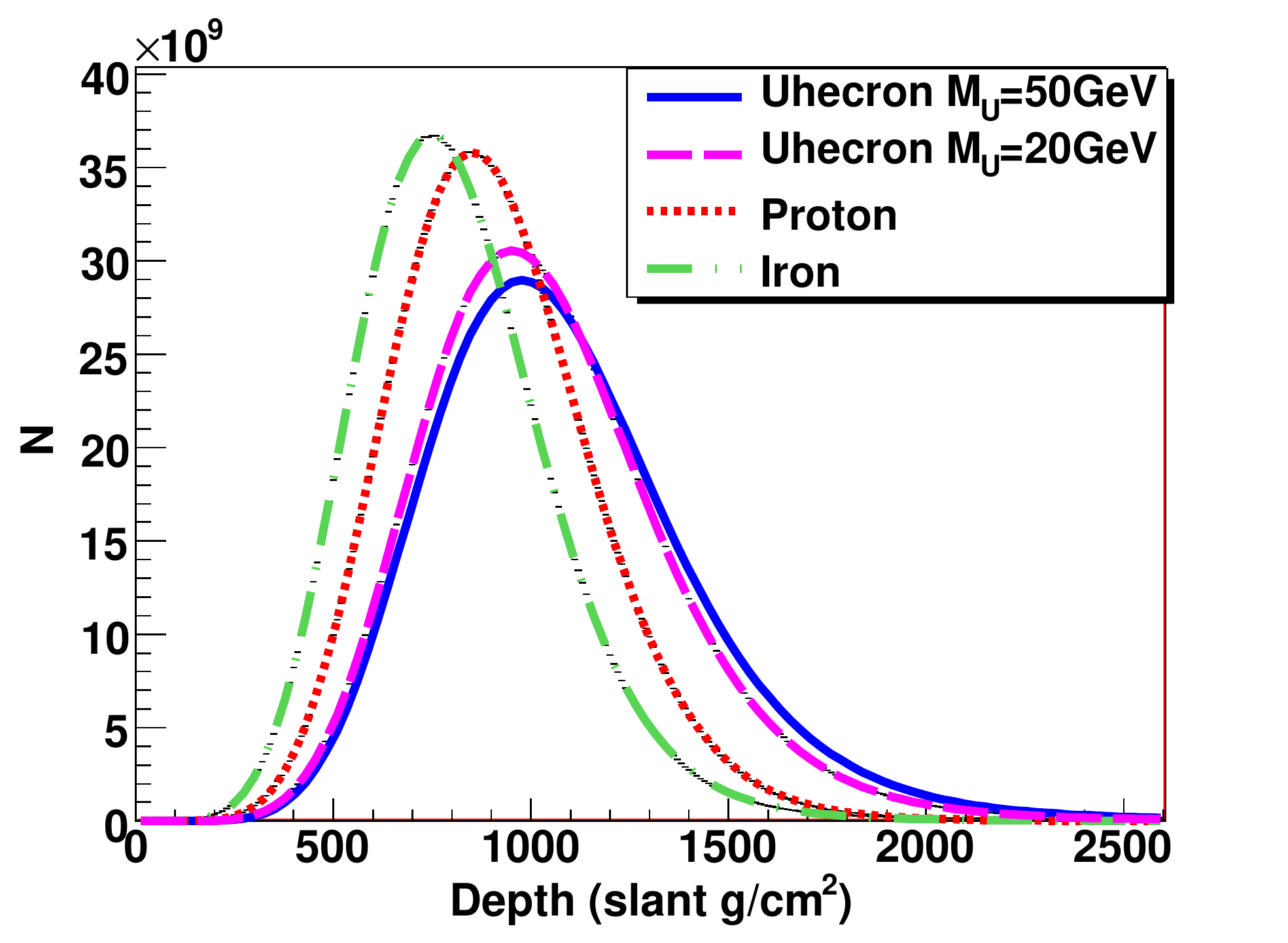}
\caption{Average longitudinal profiles 
based on 500 iron, proton and uhecron (with $m_U = 20$ and 30~GeV) 
induced showers. Primary energies are equal to 320~EeV (top) and
50~EeV (bottom). These showers were generated at a 60$^o$ zenith angle.}
\label{fig:longp}
\end{figure}

Figure~\ref{fig:longp} shows the average longitudinal profile of 320 and
50~EeV iron, proton and uhecron (with $m_U = 20$ and 50~GeV) induced 
showers based on 500 showers for each primary. As uhecrons have less energy available 
for interactions than protons, 
its shower Xmax position is deeper in the atmosphere. As the uhecron mass increases,
the available interaction energy decreases and the Xmax is deeper.
These profiles show a fit with the Gaisser-Hillas \cite{gh} function (GH) to the 
simulated data.

The Xmax position and number of particles at this position (Nmax) of each average profile 
in Figure~\ref{fig:longp} is shown in Table~\ref{tab:longp}. 
The average Xmax position of 20~GeV uhecrons is about 100~g/cm$^2$ deeper
than the one for protons for both primary energies. Although the longitudinal profile 
fluctuates, it already indicates that uhecrons resemble more protons than iron nucleus. 
For this reason we will determine ways to discriminate uhecron from protons. Our
distributions show that the same procedure will more efficiently separate them from iron.

\begin{table}
\begin{center}
\caption{Nmax and Xmax (slant depth) for shower profiles shown in Figure~\ref{fig:longp} 
and for a 30~GeV uhecron. Primary energies are 320 and 50~EeV.}
\begin{ruledtabular}
\begin{tabular}{l|cc|cc}
Energy (EeV) & \multicolumn{2}{c|}{320} & \multicolumn{2}{c}{50} \\
\hline
Particle & Nmax  & Xmax  & Nmax & Xmax \\
 & ($\times 10^{11}$) & (g/cm$^2$) & ($\times 10^{10}$) & (g/cm$^2$) \\
\hline
Iron & 2.34 & 797.1 & 3.68 & 749.4 \\
Proton & 2.23 & 897.6 & 3.58 & 852.1 \\
Uhecron (20~GeV) & 1.94 & 997.7 & 3.06 & 953.6 \\
Uhecron (30~GeV) & 1.92 & 1005.3 & 3.00 & 967.4 \\
Uhecron (50~GeV) & 1.85 & 1021.5 & 2.90 & 977.6
\label{tab:longp}
\end{tabular}
\end{ruledtabular}
\end{center}
\end{table}

\begin{figure}[h]
\includegraphics[scale=0.35]{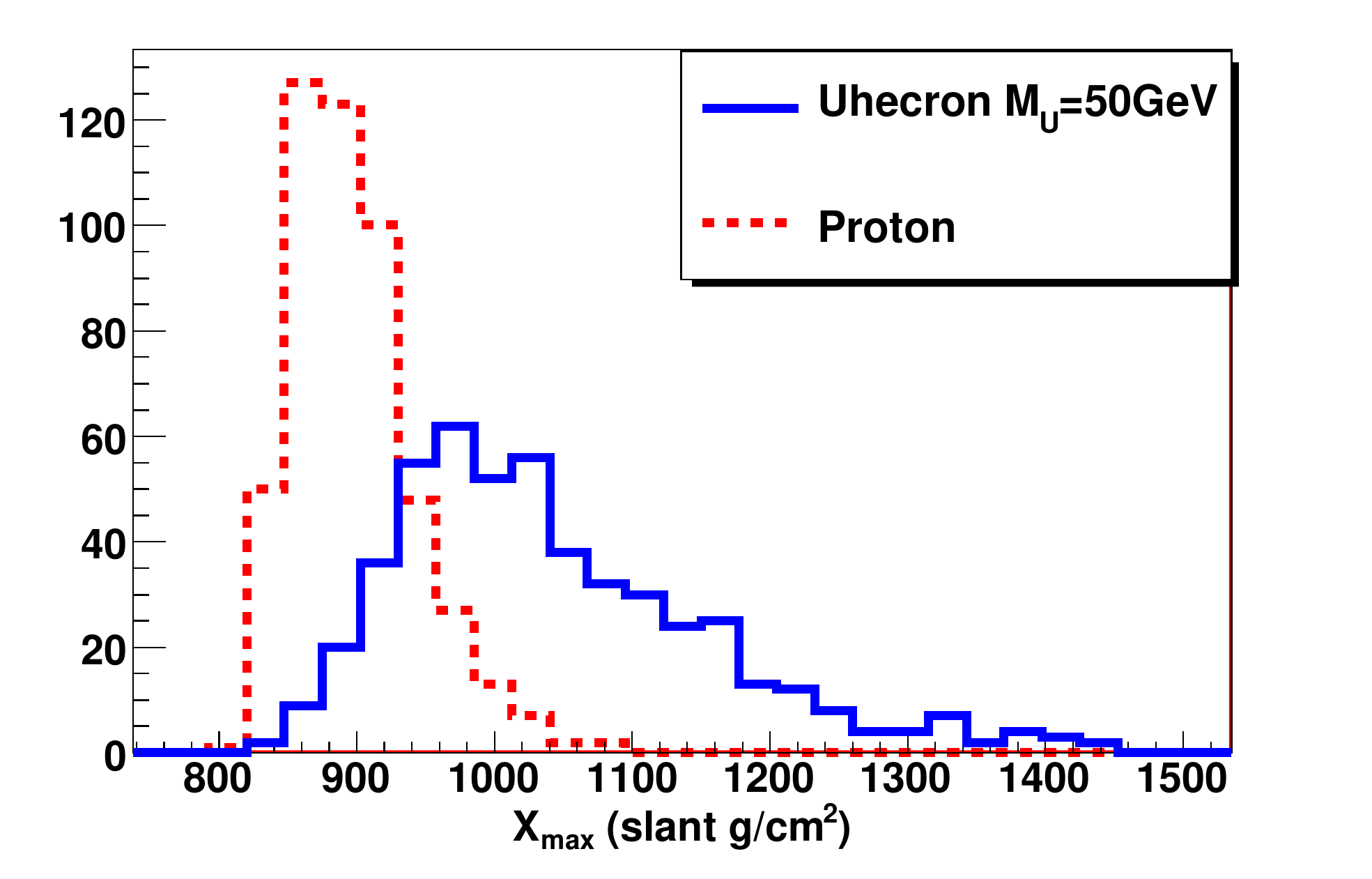}
\includegraphics[scale=0.35]{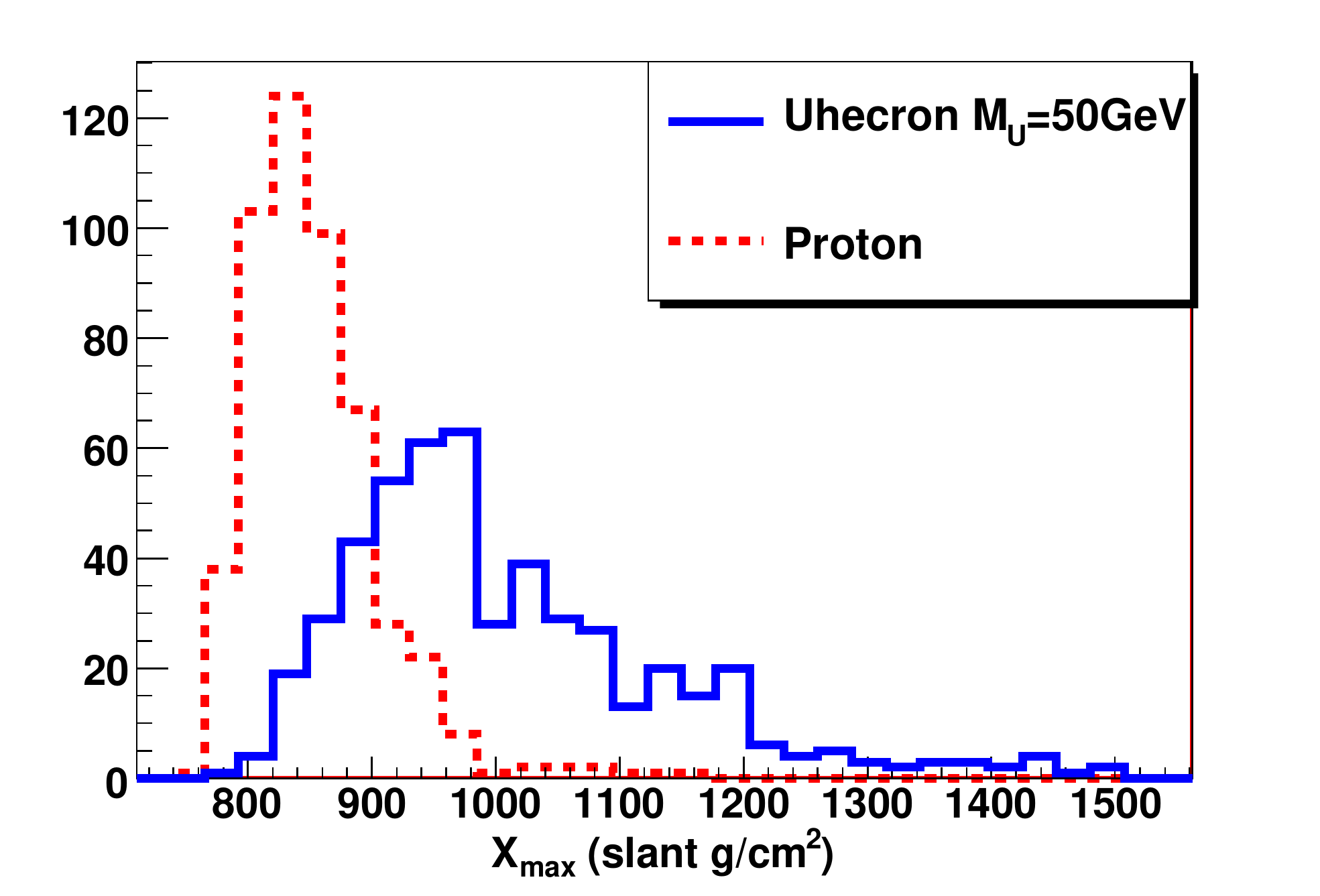}
\caption{Xmax distributions for 500 proton and 50~GeV uhecron induced showers
of 320 (top) and 50~EeV (bottom). }
\label{fig:xmaxd}
\end{figure}   

Although the average longitudinal profile gives an idea of the general
differences between proton and uhecron induced showers, the individual profile fluctuates
a lot. The Xmax distributions give a better 
idea of the fluctuation as well as the superposition among the different primaries. 
The Xmax distributions for proton and uhecron induced showers are shown in 
Figure~\ref{fig:xmaxd}. As uhecrons interact always softly, the shower fluctuations
are larger and the Xmax distribution is more spread than for protons. This feature will
help in the proton uhecron discrimination. However, as we will show in the
next section, showers with Xmax deeper than ground level are not accepted by the FD
reconstruction. This requirement ends up lowering the uhecron acceptance. 

\section{Simulation of fluorescence detection and event reconstruction}
\label{sec:fluo}

Fluorescence telescopes detect fluorescence photons emitted when 
charged particles transverse the atmosphere. 
As the air shower develops, the light emitted at different depths is collected 
by the FD photomultipliers (PMTs) and can be translated into an energy deposition
longitudinal profile.  
The integration of this profile over the full shower path is proportional to the shower
calorimetric energy. A small fraction ($\sim 10\%$) of the total shower energy is missed, 
since it is carried by neutrinos and by high energy muons which reach the ground.

After generating uhecron, proton and iron induced showers, we simulate their
detection by FDs. We use the same FD simulation as described in \cite{fluor}, which
followed the general procedure in \cite{vitfluor}. 
As our simulation aims detection of rare events, a large coverage area is needed.
For this reason we used the Pierre Auger FD parameters.
The telescope altitude is set to 1500~m above sea level, 3.8~m$^2$
aperture covering an elevation angle from 2$^o$ to 32$^o$ and using 1.5$^o$ pixel 
size PMTs. We take the telescope efficiency as 20\%.

In short, our simulation \cite{fluor}
translates the shower energy deposited at each atmospheric depth into 
production of fluorescence photons. The propagation of these photons to the detector
PMTs takes into account attenuation due to Rayleigh (molecular) and Mie (aerosol)
scattering \cite{kakimoto}. Details of fluorescence detection such as effective 
collection area,
mirror reflectivity, filter transmission, phototube quantum efficiency, noise
and background are 
included. Once the sequence of signals in each PMT is determined, we simulate the
energy reconstruction. We fold a 5$^o$ Gaussian error into the shower axis direction
and transform back the PMT signal into deposited energy. 
All effects that were taken into account in the detection simulation, are now determined
by the new reconstruction shower geometry. 

We generated 2000 showers for each primary at 3 energies (50, 100 and
320 EeV), all with a 60$^o$ zenith angle. Uhecrons with 20, 30 and 50~GeV mass
were simulated. Each of these sets were used as inputs
in the FD simulation. Each input was used 20 times, each with a different zenith
angle and core position \cite{fluor}, in order to simulate an isotropic flux. 
Overall, 40K FD events were simulated for each energy and particle.

Once the longitudinal profile was reconstructed by the FD simulation a GH function
was fit to determine the reconstructed energy. In order to cut badly
reconstructed events, basic quality cuts were applied. These are listed in 
Table~\ref{tab:qualcuts} and are always apllied in our FD event reconstruction. All 
cuts but the GH fit $\chi^2$ are typically used in
Auger analysis \cite{augan}. The GH fit $\chi^2$ was
relaxed since this fit is not as good for uhecron longitudinal profiles as
for proton's. $\Phi$ is the angle between the shower axis and the ground and is used
to minimize the Cherenkov contamination.

\begin{table}
\begin{center}
\caption{Quality requirements over FD simulated data. Events that do not
meet these requirements are cut. All but the GH $\chi^2$ are found in 
\cite{augan}. $\Phi$ is the angle between the shower axis and the ground.}
\begin{ruledtabular}
\begin{tabular}{ll}
\multicolumn{2}{c}{Quality Requirements} \\
\hline
Hit PMTs & $ > 5$ \\
Track length & $ > 200$~g/cm$^2$ \\
Zenith angle & $ < 60^o$ \\
Xmax & visible \\
$\Phi$ angle & $ < 132^o$ \\
$\chi^2$ (GH fit) & $< 50$
\label{tab:qualcuts}
\end{tabular}
\end{ruledtabular}
\end{center}
\end{table}

\begin{figure}[h]
\includegraphics[scale=0.35,angle=90]{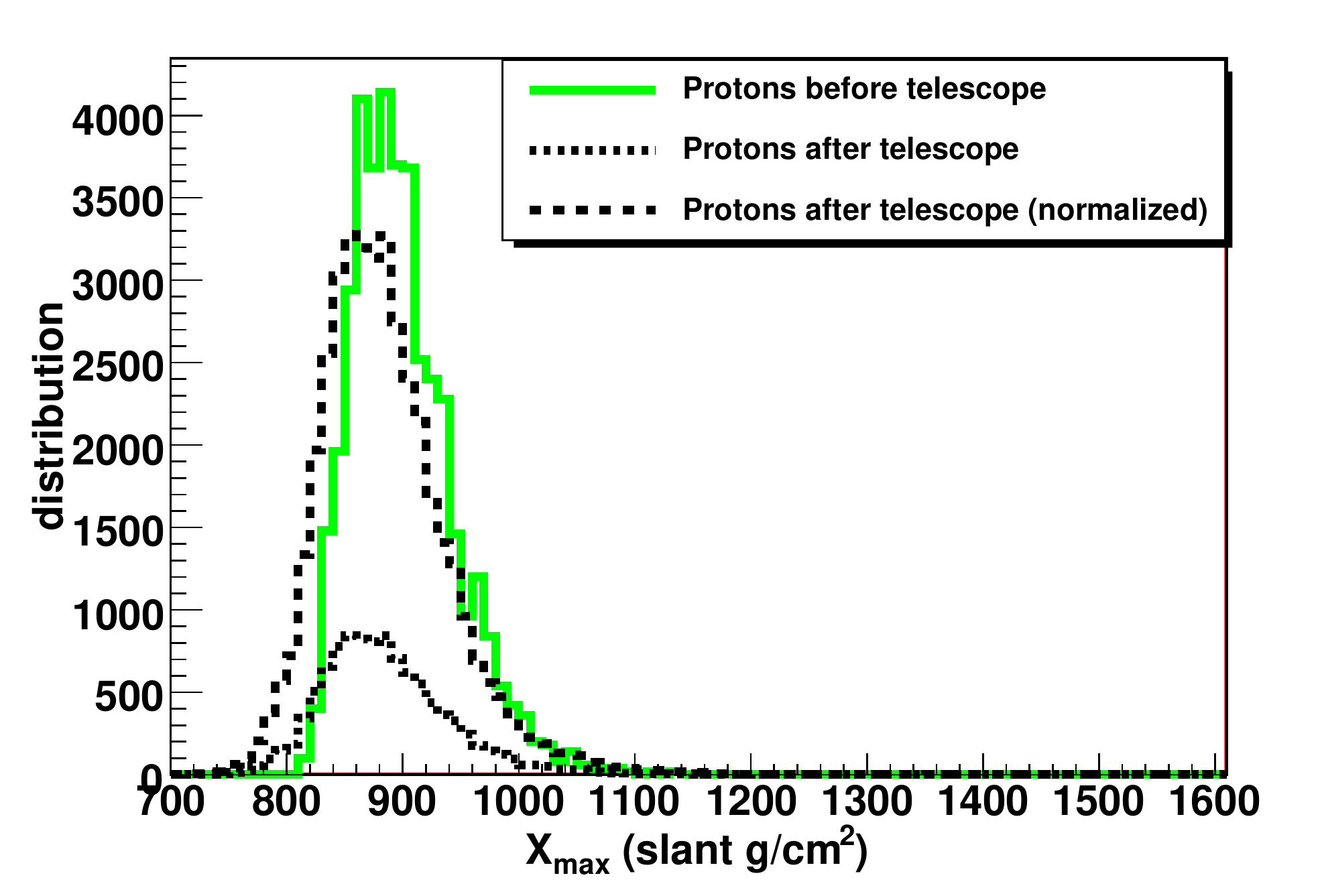}
\includegraphics[scale=0.35]{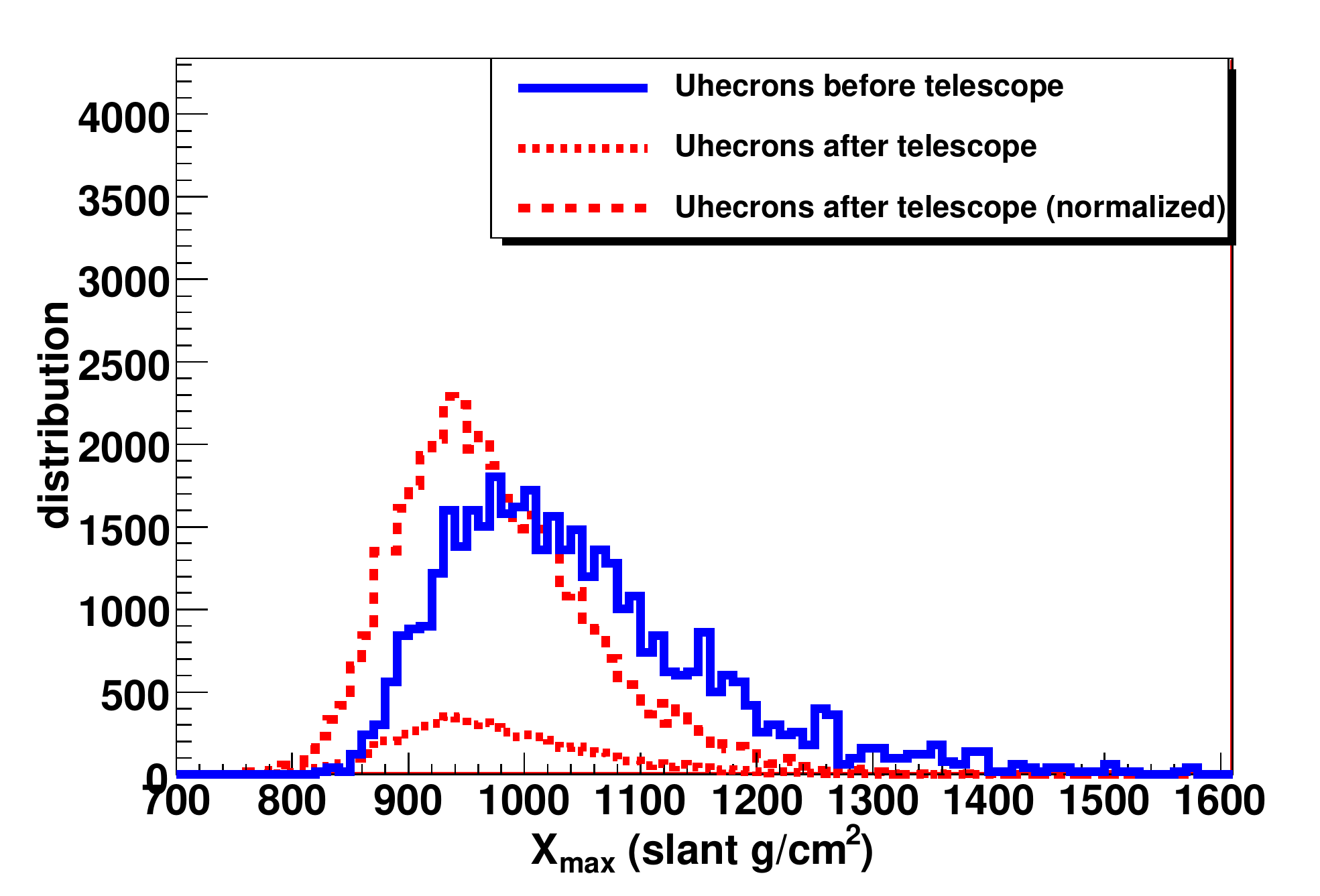}
\caption{Xmax distributions before and after FD reconstruction. Plots are
for protons (top) and 50~GeV uhecrons (bottom) with 320~EeV primary energy.}
\label{fig:xmaxfd}
\end{figure}

Detection and energy reconstruction induces errors in the reconstructed
longitudinal profile. Figure~\ref{fig:xmaxfd} compares the Xmax
distributions for protons and 50~GeV uhecrons with 320~EeV primary energy,
before and after the FD reconstruction.
For a better visualization, we also show the distribution after FD reconstruction
normalized to the number of input events.

The large reduction in the number of events comes from geometrical factors as well as
from the quality requirements. The shower detection is largely
dependent on the inclination of the shower, core position and the detector field of view 
\cite{fluor,aaron}. After the FD reconstruction, 
both proton and uhecron distributions are shifted to lower values and are 
also broader. The shift and reduction of events is weaker for protons when compared
to uhecrons. While detection
uncertainties broadens both proton and uhecron distributions, the FD acceptance
favors lower Xmax values \cite{aaron}. For this reason more uhecron events are cut 
and the 
larger Xmax side of the distribution is less accepted. This shifts the distributions
to lower Xmax values. 

Since lower energy showers have Xmax at higher altitudes, they will be less affected 
by the FD acceptance. Our distributions follow
this trend: for lower primary energies the Xmax distribution does not shift 
to lower values as much as for larger energies. Also, the reduction in the number
of events is lower. While 84.7\% (74.3\%) of 320~EeV uhecrons (protons) are cut by the 
FD reconstruction, 81.5\% (70.5\%) of 100~EeV uhecrons (protons) are cut. 

\begin{figure}[h]
\includegraphics[scale=0.35]{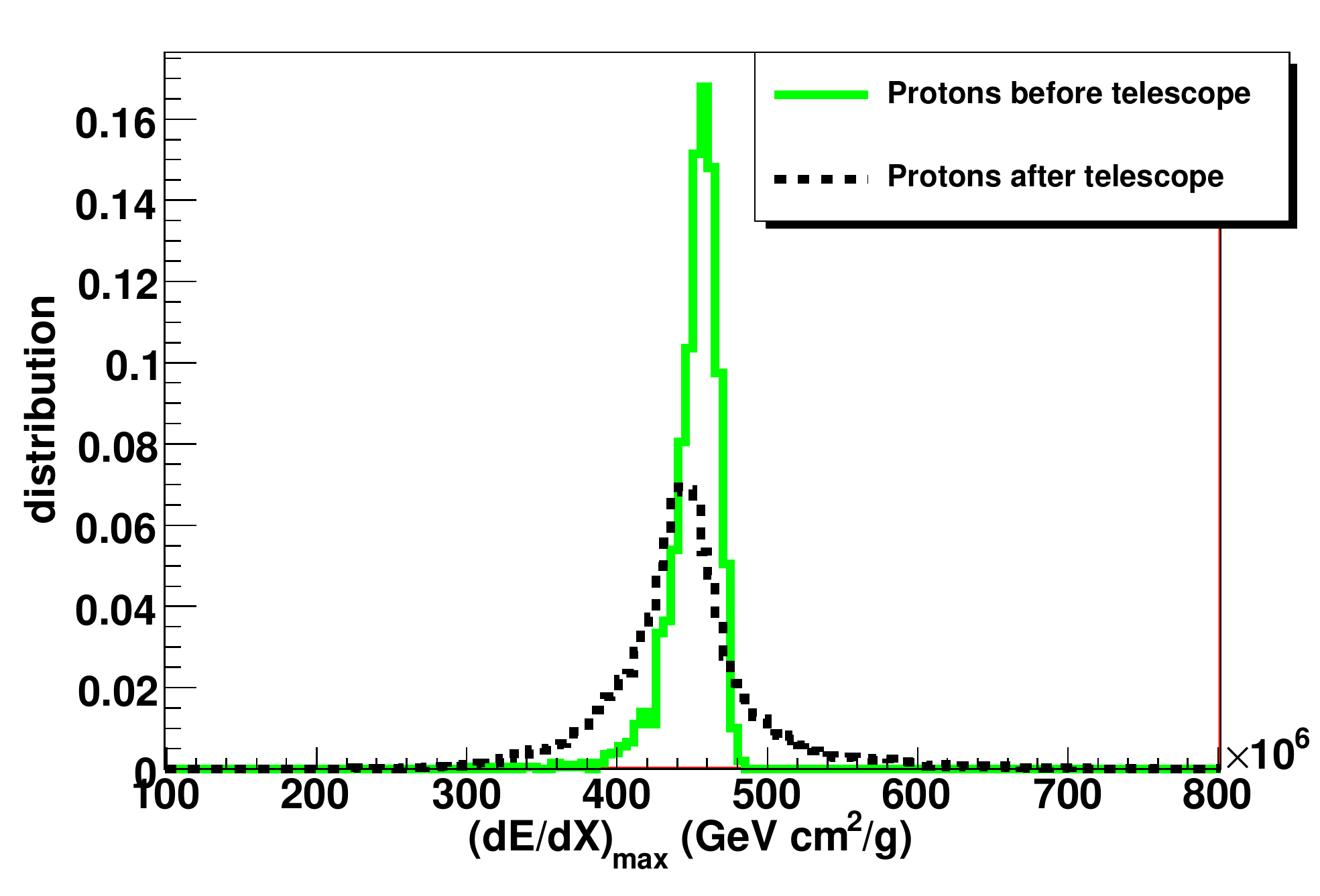}
\includegraphics[scale=0.35]{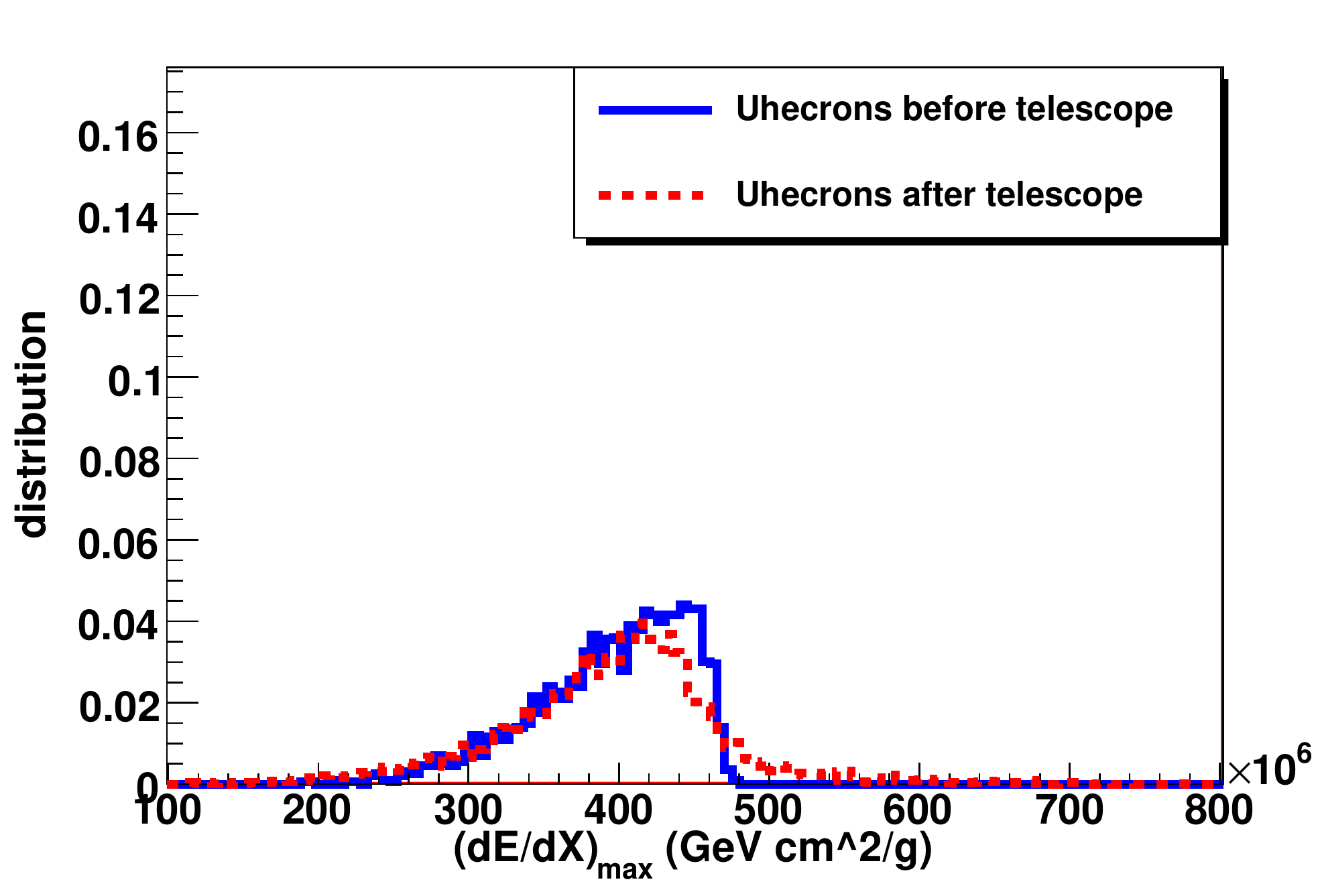}
\caption{Normalized maximum deposited energy distributions before and after FD 
reconstruction. Plots are
for 320~EeV primary energy protons (top) and 50~GeV uhecrons (bottom). The dashed
curves represent the distribution after the FD simulation normalized to the number of
events before the FD simulation.}
\label{fig:depen}
\end{figure}

Figure~\ref{fig:depen} shows the normalized maximum deposited energy distributions
$(dE/dx){\rm max}$ before and after the FD reconstruction simulation for 320~EeV 
protons and uhecrons
(with 50~GeV mass). As for the Xmax distributions, the
maximum deposited energy also shifts to lower values. While the broading of the proton 
distribution due to detection and reconstruction errors is clear on both sides,
the effect on uhecrons is not that clear, specially at lower deposited energies.
This can be explained by the inherent uhecron shower characteristics, which 
fluctuates much more than proton showers.

\begin{figure}[h]
\includegraphics[scale=0.35]{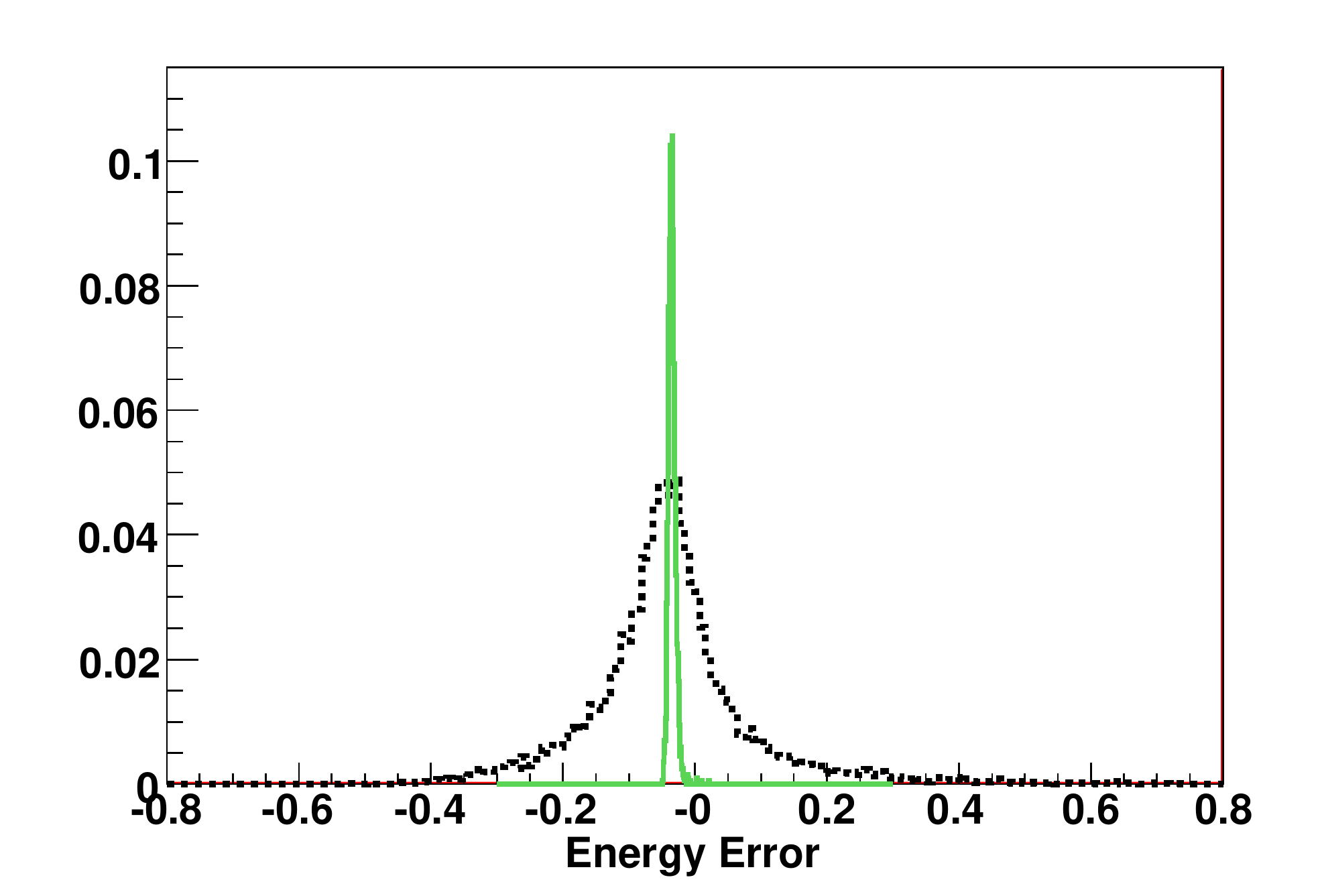}
\includegraphics[scale=0.35]{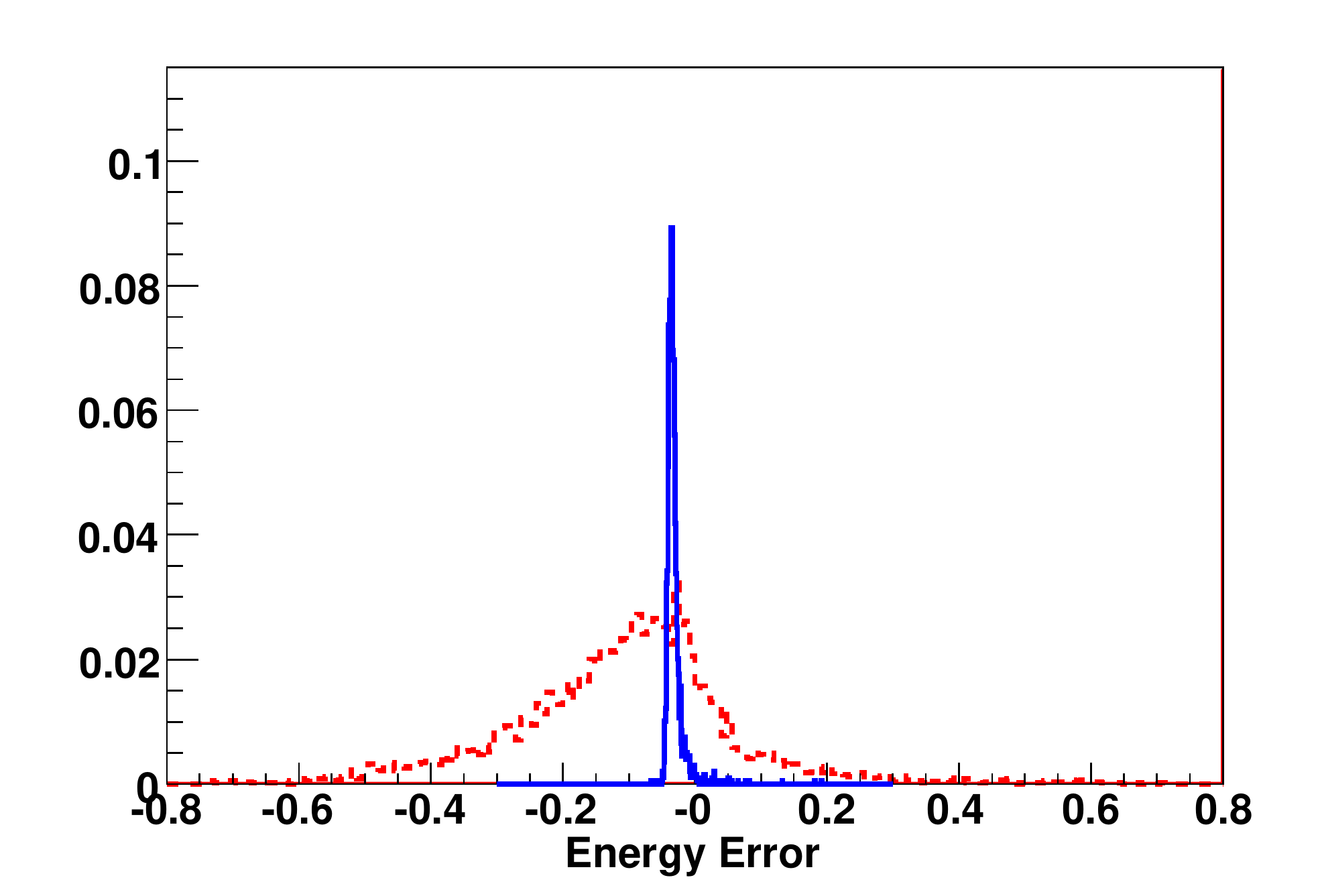}
\caption{Energy error distributions before and after FD 
reconstruction. Plots are
for 320~EeV primary energy protons (top) and 50~GeV uhecrons (bottom). The green
(proton) and blue (uhecron) narrow distributions are the errors before FD
reconstruction, while the black (proton) and red (uhecron) broad distributions
are the errors after the FD reconstruction.}
\label{fig:recen}
\end{figure}

We also compare the reconstructed energy with the primary energy. The reconstructed
energy is obtained
by adding the missing energy to the calorimetric energy. While the latter 
is determined from the integration of the energy longitudinal profile, the
missing energy is parameterized from Monte Carlo simulations.
We used the same missing energy parametrization as
determined for protons in \cite{barbosa}.  
Figure~\ref{fig:recen} shows the reconstructed energy error (given by
(E$_{\rm rec}$ - E$_{\rm primary}$)/E$_{\rm primary}$), before and
after the FD reconstruction. An energy error of about 3\% is already observed
in the reconstructed energy before the FD simulation. This error is due to the
missing energy parametrization, which was determined based
on Corsika/QGSJET \cite{cors,qgsjet} simulations and generates this error
when using AIRES/SIBYLL. Our investigation will not
be biased by this error, since we always
compare uhecrons with protons, and both are equally affected by the
missing energy parameterization.

As shown in the same Figure, the proton energy error peaks
at the same energy before and after the FD simulation. While it is symmetrically
distributed, the uhecron distribution is asymmetric. The main
reason for this, is that the GH function is not the best fit for uhecron profiles. Among
other problems it does not account for the profile tail.
In this analysis we did not attempt to find a better fit, but eventually it
can help uhecron discrimination.
As a result, uhecron showers will in average be reconstructed as
lower energy showers, with a systematic energy error around -10\%. 

\section{Uhecron -- proton discrimination}
\label{sec:analy}

As was shown in the previous section, the main characteristics of uhecron induced showers 
are larger Xmax with less particles at this position (lower
Nmax) and slower development when compared to proton
induced showers. Here we demonstrate the possibility of discriminating uhecron
from proton induced showers using FD observables. Nucleus induced showers have 
even smaller Xmax and are more easily discriminated from uhecrons.

\begin{figure}[h]
\includegraphics[scale=0.35]{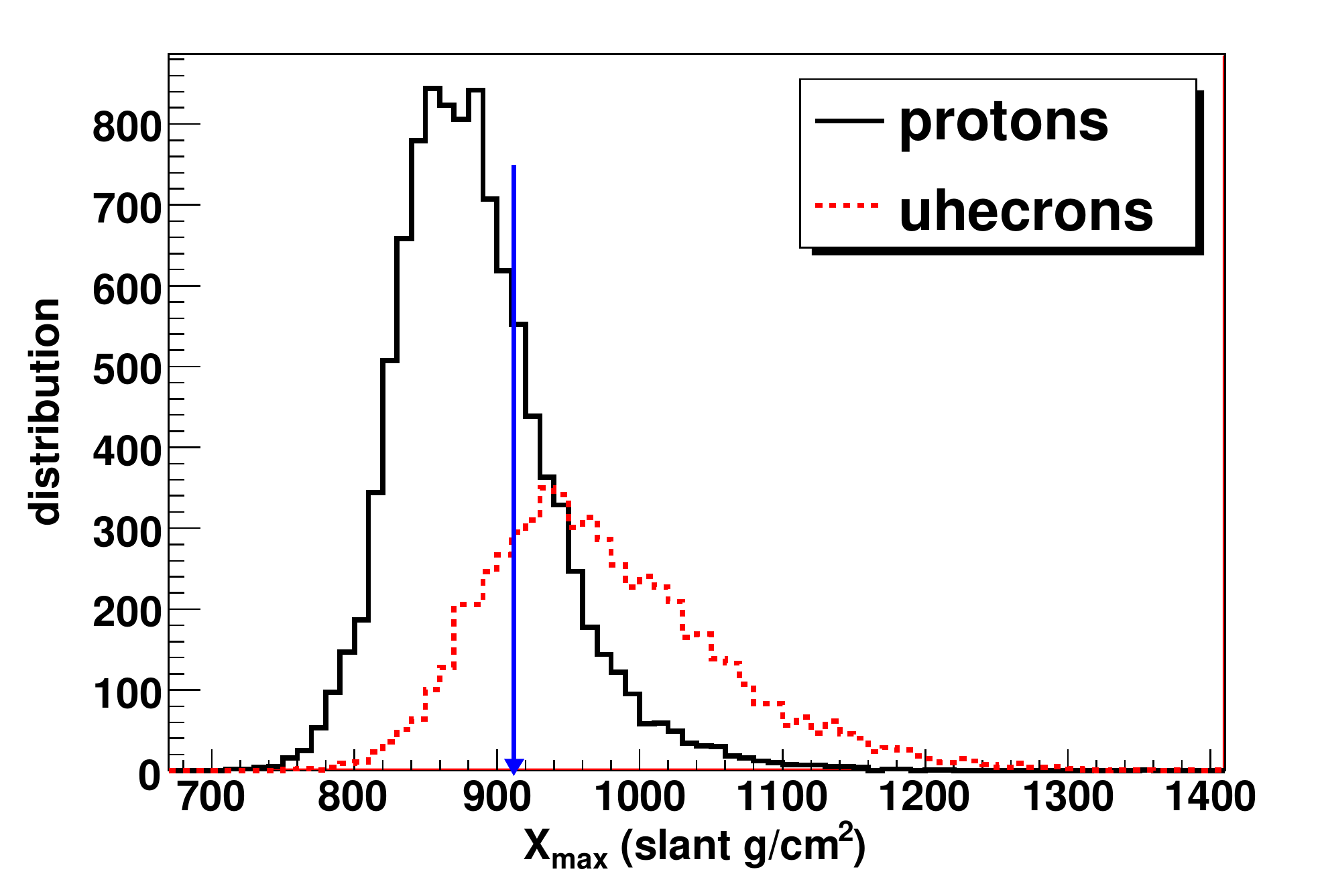}
\includegraphics[scale=0.35]{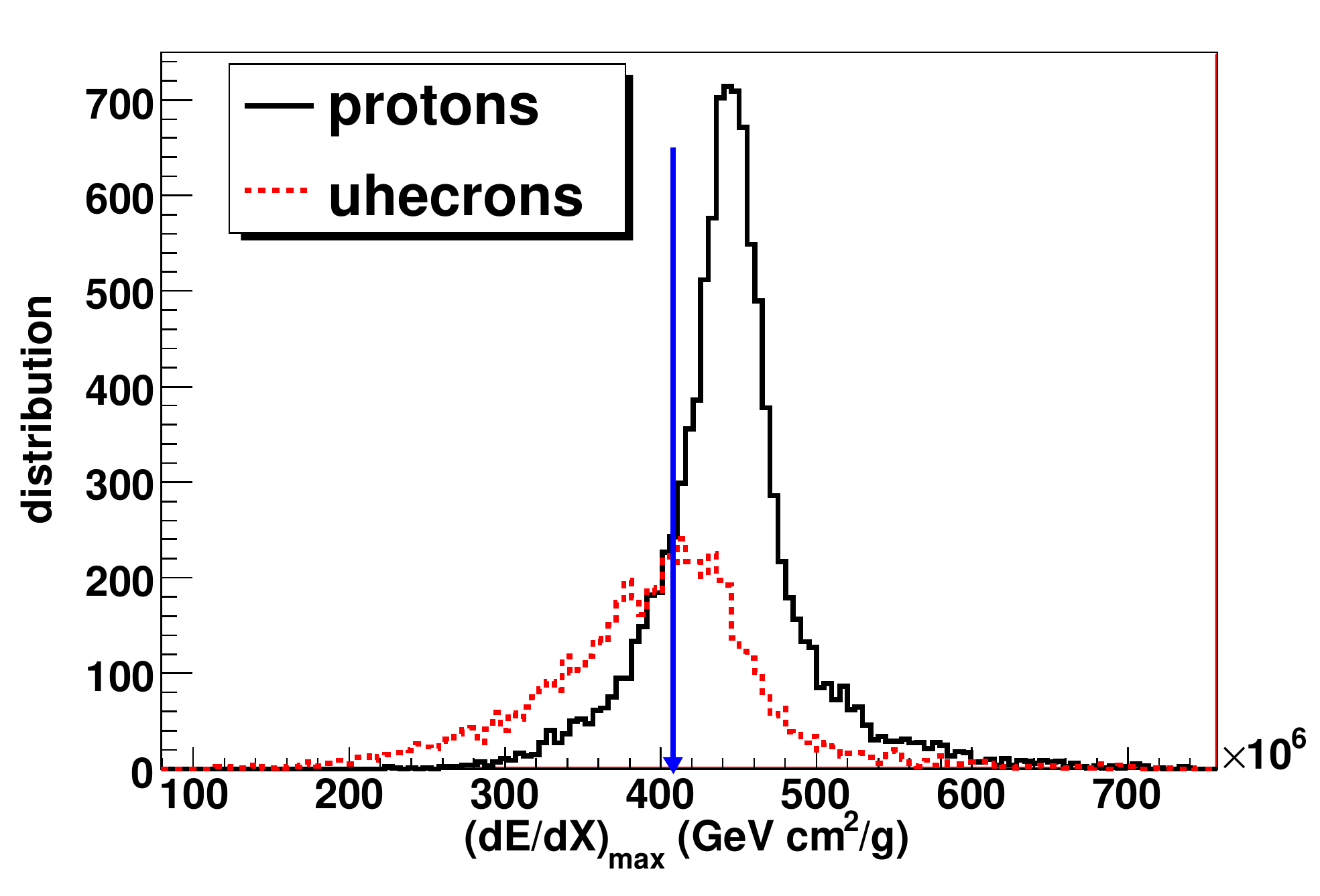}
\caption{Xmax (top) and $(dE/dx){\rm max}$ (bottom) distributions after FD 
reconstruction for 320~EeV primary energy proton and 50~GeV uhecron induced
showers. The arrows show the position of the optimized analysis cuts.}
\label{fig:xmaxcut}
\end{figure}   

Other than the Xmax and the $(dE/dx){\rm max}$
(which has the same discriminating power as Nmax),
the zenith angle $\theta_z$ and the altitude Hmax at which the first light is detected by
the FD can be used to discriminate uhecrons from protons.

It can be seen from Figures~\ref{fig:longp} and~\ref{fig:xmaxd} that uhecrons have 
deeper Xmax than protons. As a consequence a large fraction
of uhecron induced showers that come vertically into the atmosphere 
are cut by the FD reconstruction. The requirement that the shower Xmax is visible 
(see Table~\ref{tab:qualcuts}) cuts most of the vertical uhecron showers.
For this reason, uhecrons are better accepted at
larger zenith angles and a cut on low $\theta_z$ showers will be more effective
on protons.
 
As described in section~\ref{sec:uhesim}, most of the uhecron energy is 
not available for interactions. For this reason, its first
interaction point with a deposited energy larger than the FD threshold, will
be deeper in the atmosphere than the first light collected from protons. Therefore
Hmax can also be used as a discriminator.

\begin{figure}[h]
\includegraphics[scale=0.35]{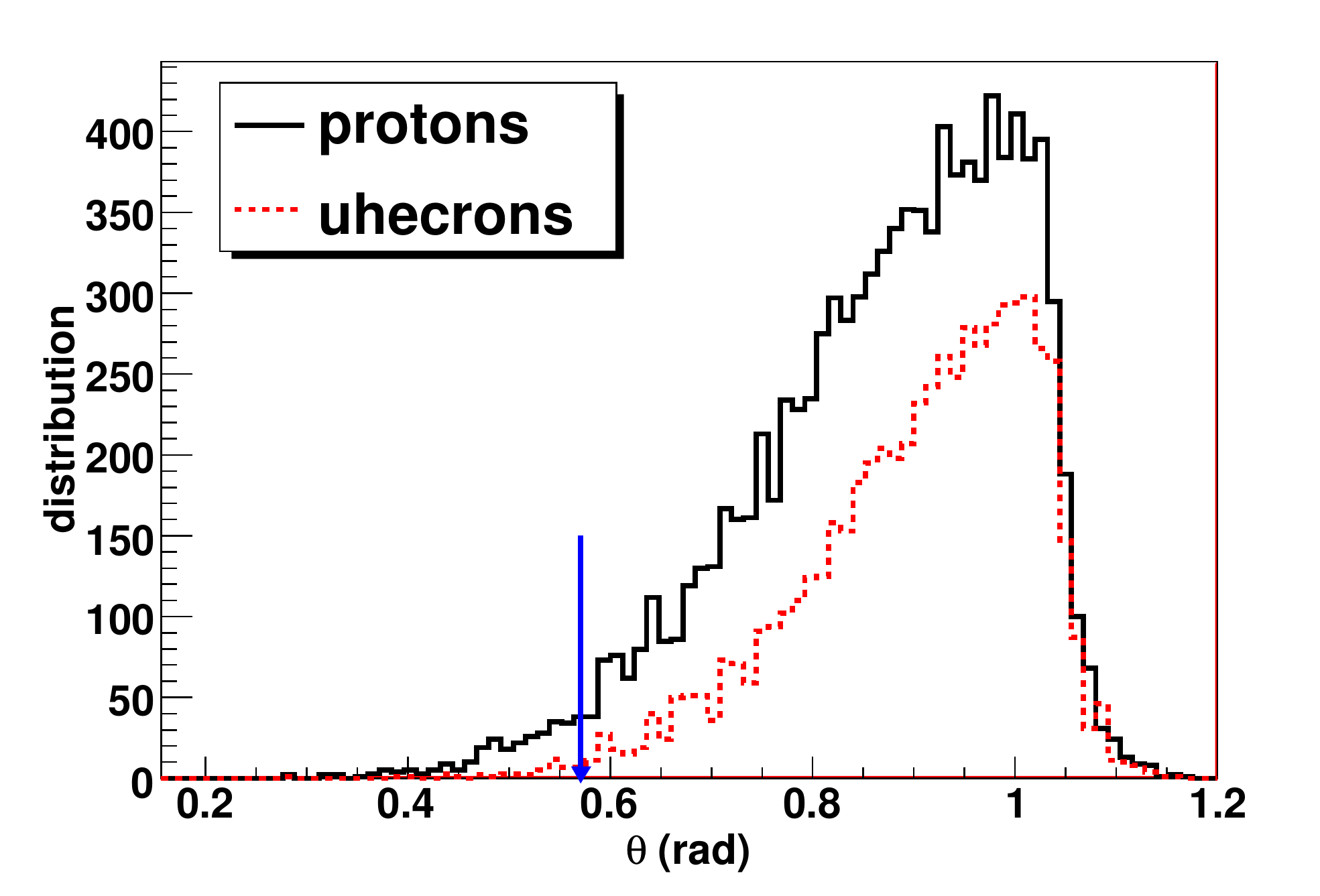}
\includegraphics[scale=0.35]{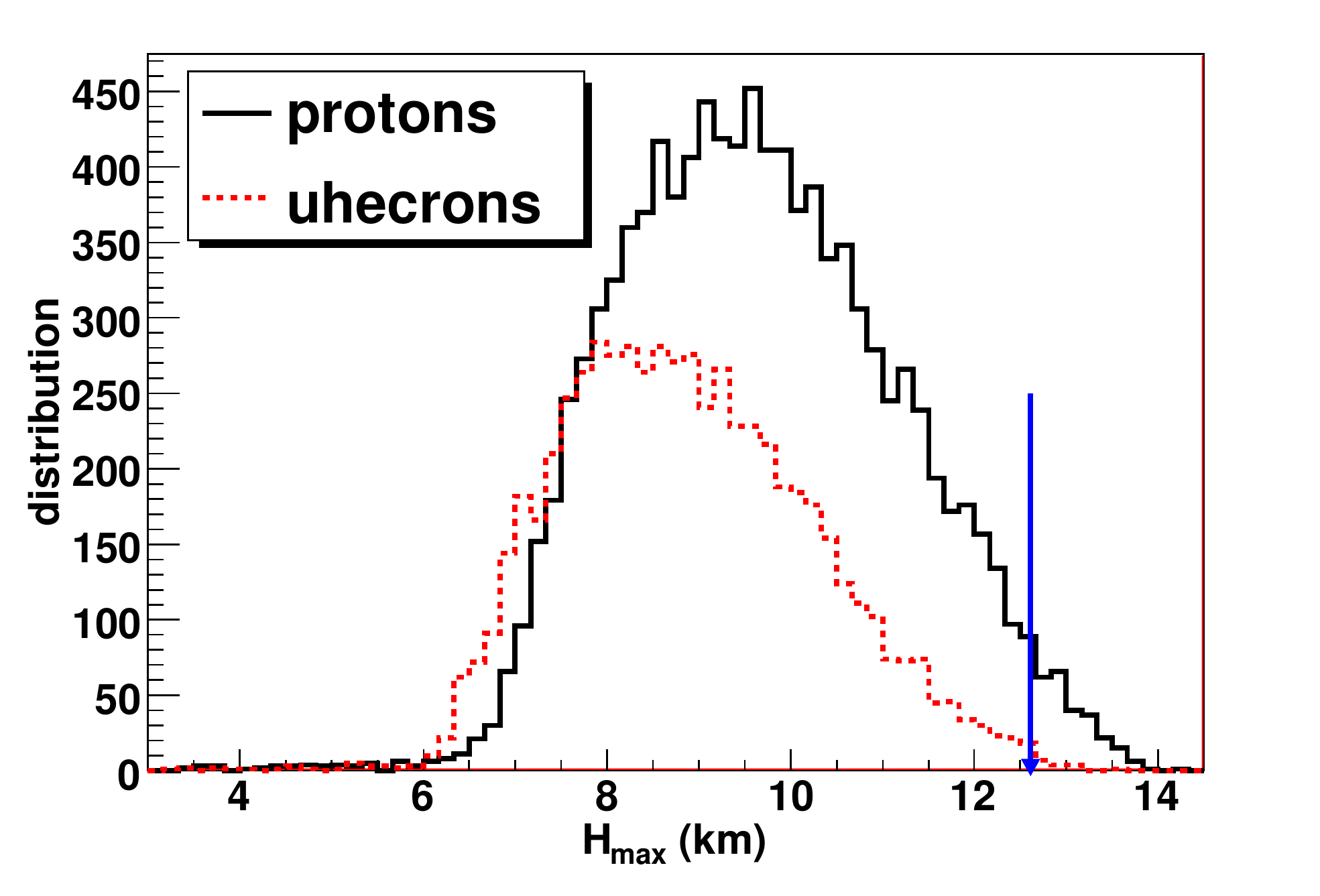}
\caption{$\theta_z$ (top) and Hmax (bottom) distributions after FD 
reconstruction for 320~EeV primary energy proton and 50~GeV uhecron induced
showers. The arrows show the position of the optimized analysis cuts.}
\label{fig:hmax}
\end{figure}

Figures \ref{fig:xmaxcut} and \ref{fig:hmax} show distributions for the observables 
used as uhecron 
discriminators. All distributions are after the FD reconstruction, for 320~EeV
showers induced by protons and by 50~GeV uhecrons. 

In order to optimize all cuts on the discriminating parameters (which from here on
we call analysis cuts), minimizing the background contamination and maximizing the 
number of uhecrons, we use the following quality factor:

\be
\label{eq:qfact}
q = \frac{N_u}{(N_u + N_p)} \times N_u^a
\ee
where $N_u$ and $N_p$ are the number of uhecrons and protons after all analysis 
cuts were applied and $a$ is a constant. Parameter $a$ sets the strenght of the cuts.

The quality factor $q$ has to be maximized as a function of the analysis cuts. 
To achieve this maximization, we scan $q$ over a fixed range for each of the
discriminating parameters Xmax, $(dE/dx){\rm max}$, Hmax and $\theta_z$.
Each combination of
cuts will yield a different $q$ factor and the maximum value will indicate
the optimized set of cuts. The arrows shown on Figures~\ref{fig:xmaxcut} and 
\ref{fig:hmax} indicate the cut values on each of these parameters. Parameter $a$ 
in Equation~\ref{eq:qfact} is set to 0.2.  

\begin{figure}[h]
\includegraphics[scale=0.35]{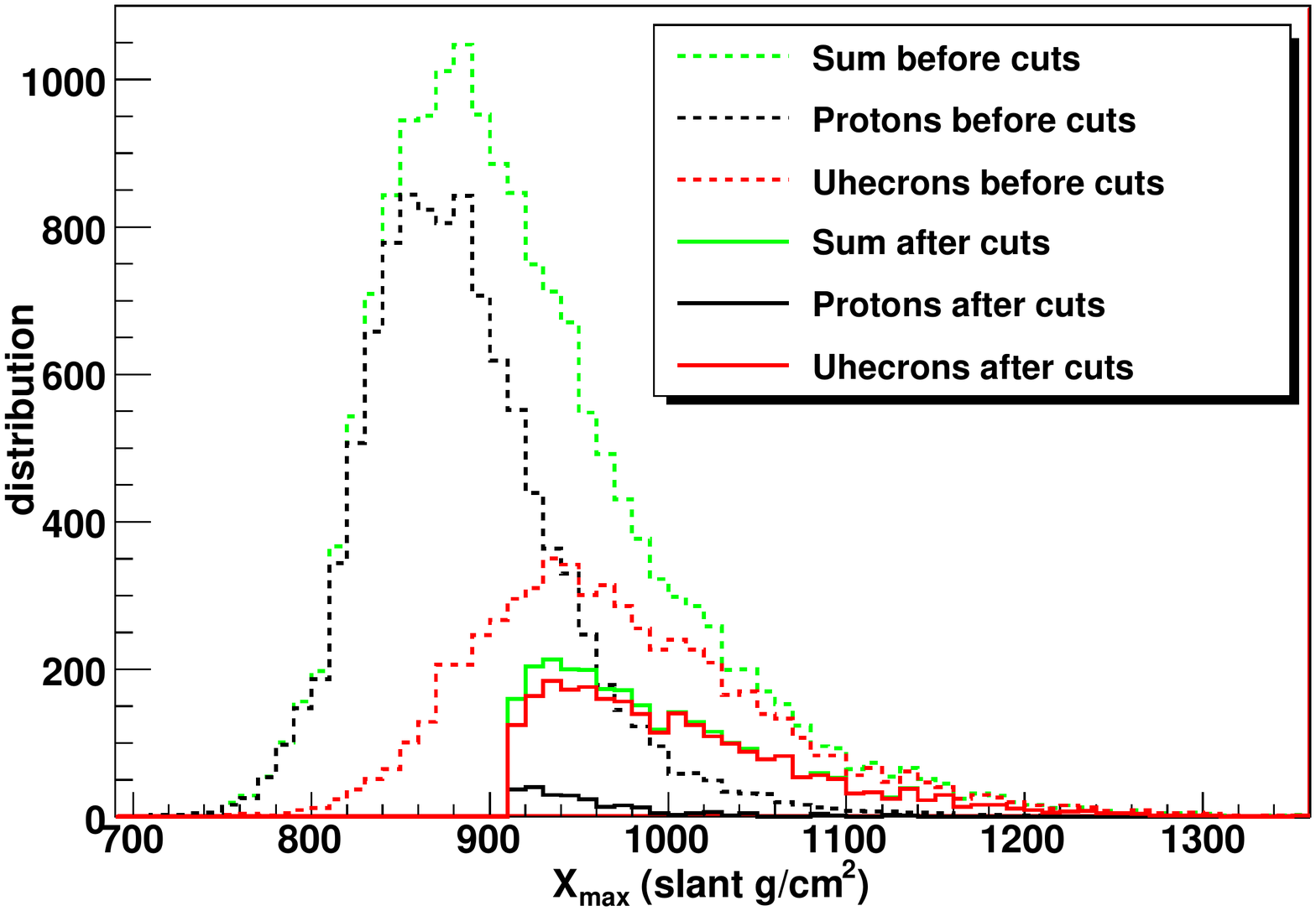}
\vspace*{-1.25cm}
\caption{Xmax distribution for protons and 50~GeV uhecrons of 320~EeV primary
energy before and after analysis cuts were applied. Cut values and fraction of events
surviving the cuts are shown in Table~\ref{tab:cut50}.}
\label{fig:xmaxaft}
\end{figure}   

Figure~\ref{fig:xmaxaft} shows the Xmax distribution for 320~EeV showers generated by
protons and by 50~GeV uhecrons before and after the analysis cuts were applied. 
It shows the effectiveness of the analysis cuts, since most of the protons are cut while
a significant uhecron fraction survives. The accepted region for each 
discriminating parameter as well as the fraction of surviving events are 
shown in Table~\ref{tab:cut50}.

\begin{table*}
\begin{center}
\begin{ruledtabular}
\begin{tabular}{cc|ccc|cccc}
$E_u$ & $E_p$ & $N_u/N_{u0}$ & $N_p/N_{p0}$ & $N_p/N_T$ & $(dE/dx){\rm max}$ &
$\theta_Z $ & Hmax & Xmax \\
 & & & & & $>$ & $>$ & $<$ & $>$ \\
\hline
320 & 320 & 0.417 & 0.022 & 0.081 & 4.08e+8 & 0.571 & 12.61 & 912.2  \\
352 & 320 & 0.402 & 0.043 & 0.152 & 5.20e+8 & 0.633 & 11.50 & 973.3 \\
108 & 100 & 0.366 & 0.039 & 0.143 & 157e+08 & 0.637 & 11.44 & 956.3 \\
54  & 50  & 0.299 & 0.016 & 0.080 & 6.64e+7 & 0.400 & 11.41 & 882.8 
\end{tabular}
\end{ruledtabular}
\caption{\label{tab:cut50} Fraction of events after analysis cuts. $E_p$ and
$E_u$ are primary energy (in EeV) of protons and 50~GeV uhecrons, respectively;
$N_p (N_{p0})$, $N_u (N_{u0})$ and $N_T$ are the number of protons; uhecrons and 
the sum of proton with uhecron induced showers after the FD
simulation and after (before) all analysis cut are applied. The last 4 columns indicate
the accepted region for each discriminating parameter after cut 
optimization, in units of GeV cm$^2$/g; rad; km and g/cm$^2$, respectively.}
\end{center}
\end{table*}

\begin{table*}
\begin{center}
\begin{ruledtabular}
\begin{tabular}{cc|ccc|cccc}
$E_u$ & $E_p$ & $N_u/N_{u0}$ & $N_p/N_{p0}$ & $N_p/N_T$ & $(dE/dx){\rm max}$ &
$\theta_Z $ & Hmax & Xmax \\
 & & & & & $>$ & $>$ & $<$ & $>$ \\
\hline
352 & 320 & 0.390 & 0.062 & 0.198 & 5.54e+8 & 0.712 & 11.41 & 961.4 \\
108 & 100 & 0.359 & 0.057 & 0.188 & 1.74e+8 & 0.616 & 10.85 & 951.7 \\
54  & 50  & 0.411 & 0.071 & 0.198 & 8.12e+7 & 0.300 & 10.90 & 922.3 
\end{tabular}
\caption{\label{tab:cut20} Same as Table~\ref{tab:cut50} but now uhecrons have 
20~GeV mass.}
\end{ruledtabular}
\end{center}
\end{table*}

As discussed at the end of the previous section, uhecrons 
will have their primary energy reconstructed
with about a 10\% error to lower values. For this reason we also compare 320 (100, 50)
~EeV proton showers with 352 (108, 54) uhecron showers, corresponding to a 10\%
(8\%) correction to the uhecron reconstructed energy. The results are shown on 
Tables~\ref{tab:cut50} and~\ref{tab:cut20}, where the first table compares
50~GeV and the latter 20~GeV uhecrons to protons. 

\begin{figure}[h]
\includegraphics[scale=0.35]{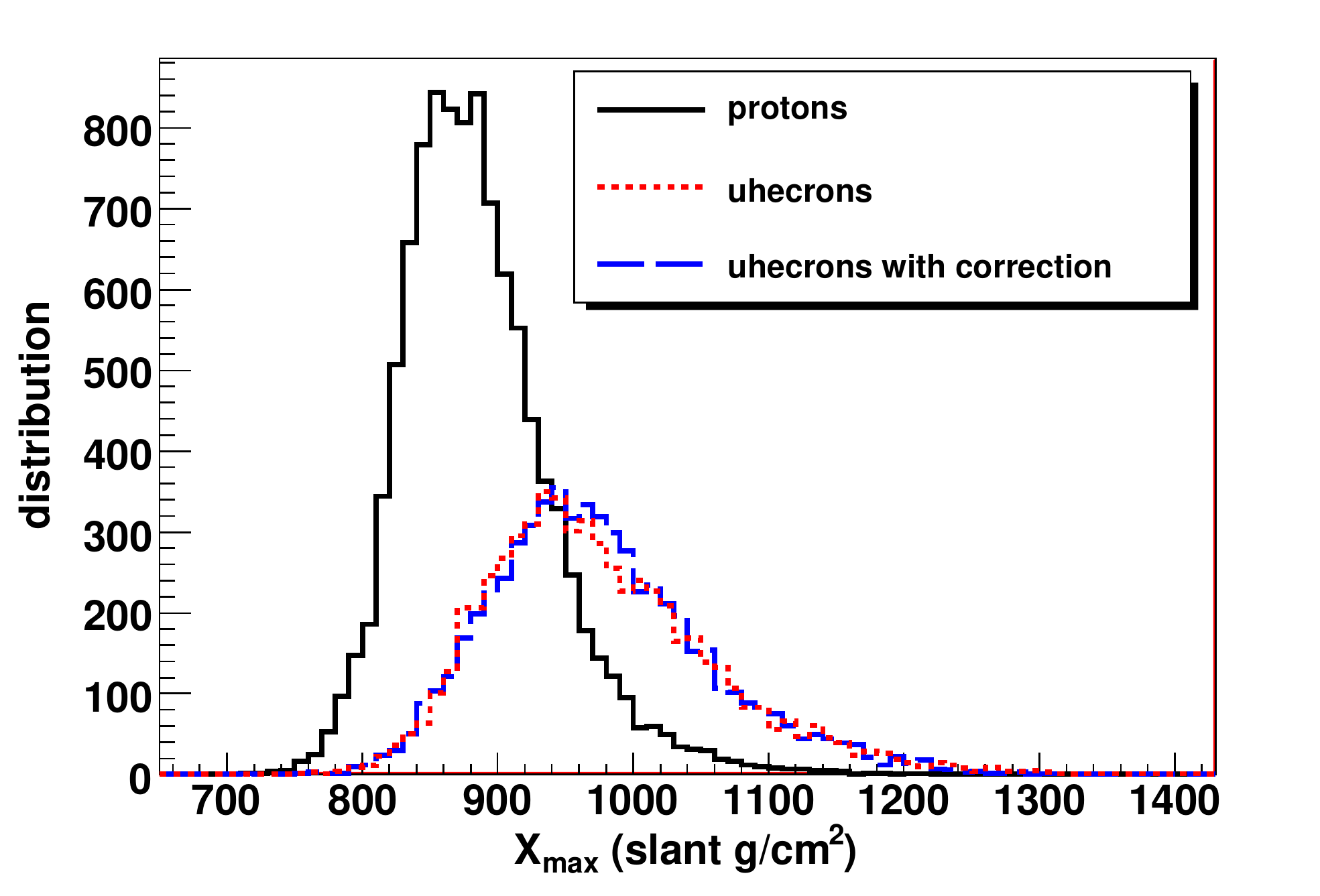}
\includegraphics[scale=0.35]{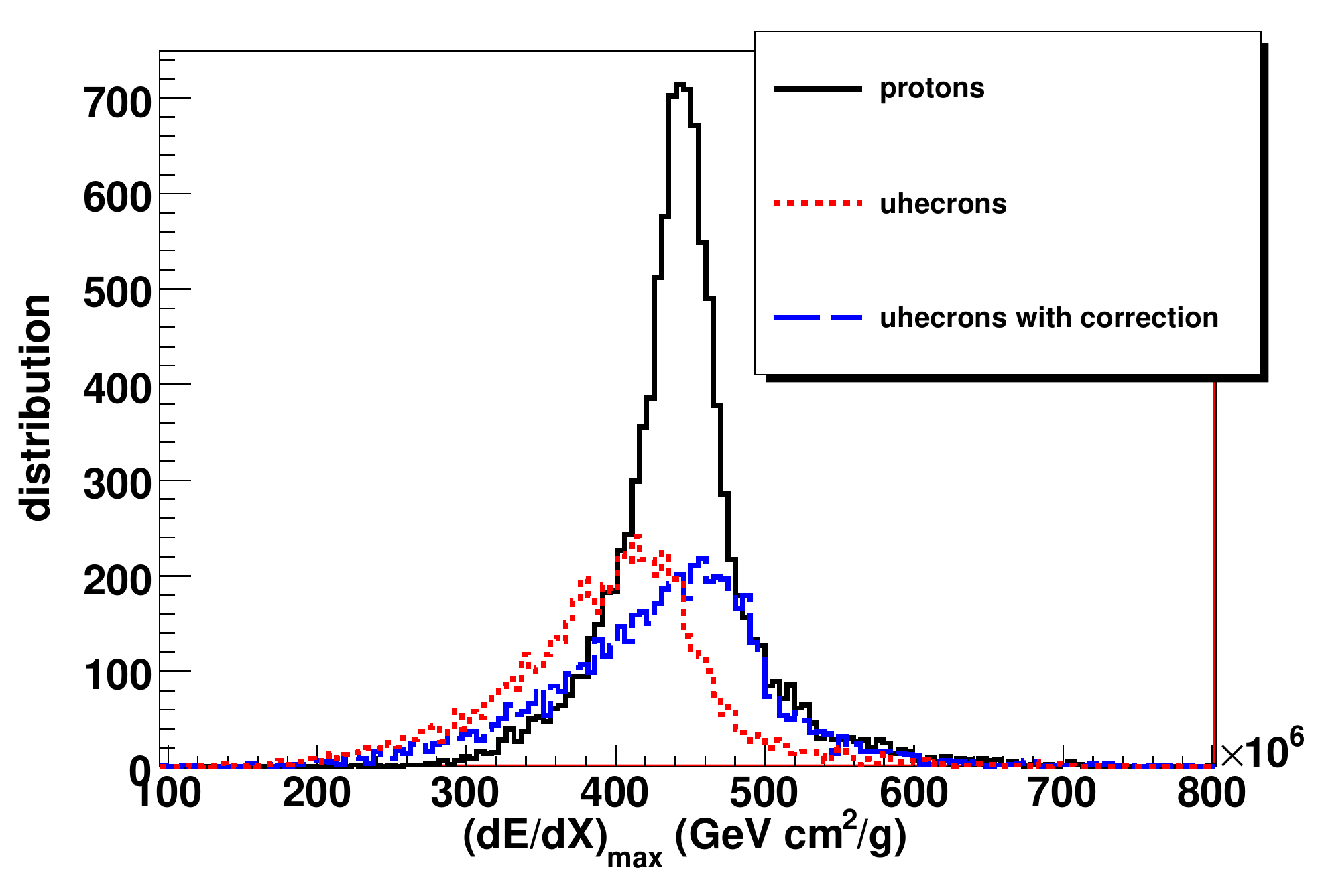}
\caption{Xmax (top) and $(dE/dx){\rm max}$ (bottom) distributions after all analysis 
cuts were applied for 320~EeV protons and 50~GeV uhecrons with 320~EeV and 352~EeV
primary energy.}
\label{fig:encorcut}
\end{figure}   

Figure~\ref{fig:encorcut} shows
both Xmax and $(dE/dx){\rm max}$ distributions 
for 320~EeV protons and 50~GeV uhecrons with 320~EeV and 352~EeV
primary energy. The Xmax distribution will not change significantly although
it shifts slightly to deeper Xmax values. However the $(dE/dx){\rm max}$
distribution changes significantly since the energy correction implies in 
a larger deposited energy. This shift in the $(dE/dx){\rm max}$ distribution
will reduce the discriminating power of this observable.

Both uhecron energy and mass are important factors for discrimination from
other primary particles. An energy increase favors deeper Xmax parameters and
therefore less FD acceptance but on the other hand enhances the intrinsic 
differentiating shower characteristics in relation to protons. Lower uhecron
masses approximate their shower intrinsic characteristics to proton showers,
but increases the uhecron FD acceptance. 
As shown in Tables~\ref{tab:cut50} and~\ref{tab:cut20}, it is possible
to greatly reduce the proton contamination using Xmax, $(dE/dx){\rm max}$, Hmax
and $\theta_z$ as discriminating parameters. The proton contamination in the final
event sample is at maximum 15\% for 50~GeV uhecrons and below 20\% for 20~GeV
uhecrons.

\subsection{Uhecron -- proton flux ratio}

Up to now we have considered the same number of input uhecron and proton induced
showers into the FD simulation. This is equivalent to an equal flux of protons and
uhecrons arriving at the Earth. However, considering the latest UHECR flux
measurements \cite{auflx,hrflx} a much lower uhecron flux has to be considered 
at least up to energies around the expected GZK cutoff. Beyond this point,
a nucleon or nucleus flux is not expected. Events at energies beyond the GZK
cutoff might indicate new physics. 

\begin{table}
\begin{center}
\begin{ruledtabular}
\begin{tabular}{cc|ccc|cc}
$E_u$ & $E_p$ & $\phi_u/\phi_p$ & $a$ & $N_u/N_{u0}$ & $N_p/N_{p0}$ & $N_p/N_T$ \\
\hline
352 & 320 & 0.1 & 0.6 & 0.139 & 0.0023 & 0.219 \\
352 & 320 & 0.05 & 0.5 & 0.102& 0.0006& 0.161 \\
\hline
108 & 100 & 0.05 & 0.6 & 0.082& 0.0006& 0.184 \\
108 & 100 & 0.05 & 0.7 & 0.162& 0.0044& 0.458 \\
\hline
54  & 50  & 0.05 & 0.6 & 0.158& 0.0015& 0.234 \\
54  & 50  & 0.01 & 0.8 & 0.081& 0.0001& 0.156 \\
\end{tabular}
\caption{\label{tab:perc} Fraction of events after analysis cuts for protons 
and 50~GeV uhecrons. 
$\phi_u/\phi_p$ is input ratio of uhecron to proton induced showers into the FD
simulation. $a$ is a parameter in Equation~\ref{eq:qfact} and all other 
parameters are described in Table~\ref{tab:cut50}.}
\end{ruledtabular}
\end{center}
\end{table}

Here we redo the analysis described in section~\ref{sec:analy}, but reducing the 
uhecron flux $\phi_u$ to 10, 5 and 1\% relative to the proton flux $\phi_p$. 
We analyse a 1\% uhecron fraction for a 50~EeV primary energy (54~EeV for a uhecron
shower) since at lower energies uhecrons might be present as a small fraction of the
flux. Even in this 
scenario it is possible to discriminate uhecrons from protons. 
We summarize our results in Table~\ref{tab:perc}. In order to enhance the final 
number of uhecron events,
we use larger $a$ parameter values (see Equation~\ref{eq:qfact}). After all 
analysis cuts are applied, the proton contamination in the final sample, for
a 1\% uhecron flux is around 
16\%. As we will discuss in the last section, our results indicate the feasibility 
of discriminating uhecrons from protons, even with a much smaller uhecron flux.



\subsection{Sample independence test}

In order to check our uhecron analysis and the discriminating power of the
Xmax, $(dE/dx){\rm max}$, $\theta{\rm max}$ and Hmax observables, we applied the
same analysis cuts as described above to a new set of data. This new set of data
uses the same 2000 showers generated from our shower simulation, for each primary
particle (where 3 different uhecron masses -- 20, 30 and 50~GeV -- were assumed)
and for each different primary energy (50, 100 and 320 EeV) and input it with
different geometry \cite{fluor} than the original analysis to the 
FD simulation. The analysis 
cuts applied to this new simulated data set were the ones determined in the original 
analysis.

\begin{table*}
\begin{center}
\begin{ruledtabular}
\begin{tabular}{cc|ccc|cccc}
$E_u$ & $E_p$ & $N_u/N_{u0}$ & $N_p/N_{p0}$ & $N_p/N_T$ & $(dE/dx){\rm max}$ &
$\theta_Z $ & Hmax & Xmax \\
 & & & & & $>$ & $>$ & $<$ & $>$ \\
\hline
352& 320& 0.395& 0.044& 0.164& 5.20e+08& 0.633& 11.50& 973.3\\ 
320& 320& 0.414& 0.025& 0.090& 4.08e+08& 0.571& 12.61& 912.2\\
108& 100& 0.358& 0.042& 0.156& 1.57e+08& 0.637& 11.44& 956.3\\
54&  50&  0.306& 0.019& 0.090& 6.64e+07& 0.400& 11.41& 882.8
\end{tabular}
\end{ruledtabular}
\caption{\label{tab:ind50} Same as Table~\ref{tab:cut50}. Cuts are now applied to new
data set. Accepted region for each discriminating parameter was defined from
original data set. Uhecron mass was set to 50~GeV. Results with original and with new
data set are compatible.}
\end{center}
\end{table*}

\begin{table*}
\begin{center}
\begin{ruledtabular}
\begin{tabular}{cc|ccc|cccc}
$E_u$ & $E_p$ & $N_u/N_{u0}$ & $N_p/N_{p0}$ & $N_p/N_T$ & $(dE/dx){\rm max}$ &
$\theta_Z $ & Hmax & Xmax \\
 & & & & & $>$ & $>$ & $<$ & $>$ \\
\hline
352& 320& 0.390& 0.062& 0.198& 5.54e+08& 0.712& 11.41& 961.4\\ 
108& 100& 0.359& 0.057& 0.188& 1.74e+08& 0.616& 10.85& 951.7\\
54&  50&  0.411& 0.071& 0.198& 8.12e+07& 0.300& 10.90& 922.3
\end{tabular}
\caption{\label{tab:ind20} Same as Table~\ref{tab:ind50} but now uhecrons have 
20~GeV mass.}
\end{ruledtabular}
\end{center}
\end{table*}

We obtain similar results as in the original analysis. 
The analysis cuts have the same discriminating power. Tables~\ref{tab:ind50}
and~\ref{tab:ind20} summarize the analysis results for this new data set
(for 50~GeV and 20~GeV uhecrons respectively). As can be seen the results
are compatible with the ones in Tables~\ref{tab:cut50} and~\ref{tab:cut20}.

\section{Uhecron -- photon comparison}


Photon showers develop deeper in the atmosphere than proton showers.
For this reason it resembles more uhecron than proton induced showers.
However it is important to note that the competition 
among uhecron and photons is not realistic. At these energies, both photons and 
uhecrons are proposed in beyond the standard model of particle physic scenarios. 
These either propose ultra high energy photons or exotic hadronic particles.
It is already known that the photon fraction of the UHECR flux is very small,
which constrain many top-down models~\cite{gunter}. In these models, ultra
high energy photons would be produced from exotic heavy particle decay.
Auger results limit \cite{augphoton} the photon fraction of the UHECR flux to 2\% \
(5.1\% and 31\%) of
the total flux above $1\times$ ($2\times$ and $4\times$) 10$^{19}$~eV with
95\% CL.  

It is also important to note that photon induced showers develop differently from
hadronic induced showers. This difference is maily due to smaller particle multiplicity
in the eletromagnetic cascade when compared to a hadronic cascade. As a consequence
the photon shower Xmax is in average deeper than the proton Xmax. Also, the number of 
muons in hadronic showers is greater than in photon showers, due to charged
pion decay. For this reason, a FD uhecron -- photon discrimination can be greatly 
enhanced by ground detector information.

\begin{figure}[h]
\includegraphics[scale=0.35]{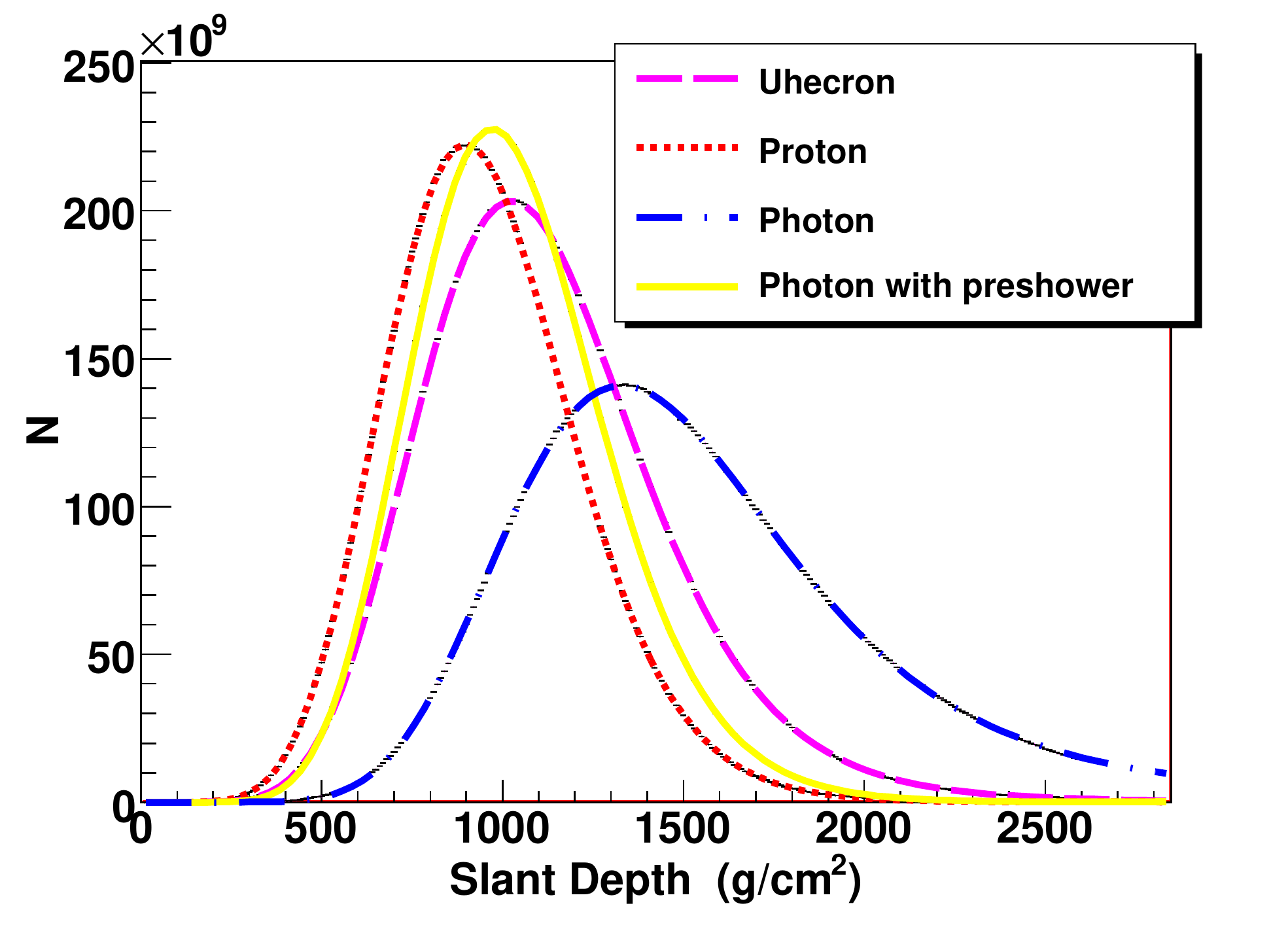}
\caption{\label{fig:pres} Longitudinal profile for photon induced showers with 
and without the preshower effect. Proton and 50~GeV uhecrons are also shown.
Proton and photon showers have 320~EeV primary energies while uhecron
has 352~EeV. All profiles are before FD reconstruction.}
\end{figure}   

Another important effect to be taken into account is due to photon interactions
with the Earth geomagnetic fields. As a consequence they preshower 
before entering into the atmosphere. In Figure~\ref{fig:pres} we compare
proton, uhecron (with 50~GeV mass) and photon longitudinal profiles for
320~EeV induced showers. Photon profiles are shown with and without
preshower. The effect seen in the longitudinal profile will be present
in the Xmax distribution as well. The preshower effect changes both
photon longitudinal and Xmax distributions in a way that it will 
resemble more uhecron showers than if this effect was not present.
However as mentioned above the hadronic characteristics of
uhecron showers might allow for their separation \cite{tocome}.

\section{Discussion and conclusion}

We have shown that UHECR experiments, such as the Pierre Auger Observatory, have the
potential to detect exotic massive hadrons. Also that it is possible
to discriminate them from nucleon induced showers. 

Although both proton and uhecrons produce hadronic showers, uhecron characteristics 
will allow discrimination from protons. As the uhecron mass increases, its induced shower
develop more slowly and fluctuates more. These characteristics allow to 
better distinguish heavier uhecron from proton showers, although it is also possible
to distinguish lighter uhecrons. While the proton contamination in the final 
simulated data sample, after all analysis cuts are applied, is at maximum 15\% for
50~GeV uhecrons it is around 20\% for 20~GeV uhecrons. 

We have also shown the effects of fluorescence detection and event reconstruction.
FD requirements can exclude uhecron showers that are naturally better discriminated from
protons. However even after FD detection and event reconstruction it is possible to
separate showers induced by these two primaries. We have shown that FD observables
such as Xmax, $(dE/dx){\rm max}$, $\theta{\rm max}$ and Hmax, are good discriminators.

Although we have no prediction for the ratio between proton and uhecron induced 
showers, we have shown that the uhecron flux can be as small as 1\% of the total
flux and still be discriminated from protons. At lower primary energies, standard
model particles should dominate the UHECR spectrum, whereas at energies beyond the
GZK cutoff it is possible to have a larger exotic flux.

It is important to note that the search for beyond standard model particles is
complementary to accelerator searches. It depends on an assumed
prior model. If the Large Hadron Collider (LHC) has indication of a heavy 
gluino~\cite{raby,rabycdf}, UHECR
telescopes can look for it in a complementary way. Or vice-versa, one can find
uhecron candidates among UHECR and depending on LHC results
investigate its identity. Heavy gluino \cite{raby,rabycdf} and strongly interacting 
Wimpless particles \cite{feng} are examples of uhecrons. The current allowed
heavy gluino mass window~\cite{rabycdf} (25 to 35~GeV) is within the uhecron
mass limits that allow separation from proton or nuclei background. 

We have also shown that our method for uhecron detection and background reduction
is independent from our simulated data. After our analysis method was determined,
we applied it to a new data set. The new results show that our discriminating parameters
have the same power as when applied to the original data set.

We also have compared uhecron to ultra high energy photon induced showers.
Although it is not expected to have both these particles as UHECR primaries,
the differences between a hadronic and photon induced shower should allow for their
discrimination~\cite{tocome}. Ground detectors should improve this discrimination.


As the uhecron flux at energies around the GZK cutoff has no reason to be large,
the construction of the northern Auger site will definetely improve the uhecron
detection probability.

\begin{acknowledgments} 
The authors thank Vitor de Souza for his useful comments. They also
acknowledge the support of the State of S\~{a}o Paulo 
Research Foundation (FAPESP) and I.A. is also supported by the Brazilian National 
Counsel for Technological and Scientific Development (CNPq).
\end{acknowledgments}

\end{document}